\input{aipcheck}

\documentclass[
    ,final            
  ]
  {aipproc}

\layoutstyle{6x9}

\usepackage{epsfig}
\usepackage[fleqn,sumlimits,intlimits]{amsmath}
\usepackage{latexsym,amssymb}
\usepackage[english]{babel}
\usepackage{verbatim}
\usepackage{color}
\usepackage{varioref}

\newcommand{\bea}{\begin{eqnarray}}
\newcommand{\eea}{\end{eqnarray}}
\newcommand{\beq}{\begin{equation}}
\newcommand{\eeq}{\end{equation}}

\newcommand{\bal}{\begin{align}}
\newcommand{\eal}{\end{align}}
\newcommand{\bit}{\begin{itemize}}
\newcommand{\eit}{\end{itemize}}

\newcommand{\rar}{\rightarrow}

\newcommand{\abs}[1]{\vert #1\vert}
\newcommand{\avg}[1]{\left\langle #1 \right\rangle}

\newcommand{\half}{{\frac{1}{2}}}

\newcommand{\dH}{{d_H}}
\newcommand{\dS}{{d_s}}

\newcommand{\meas}{{\mu(\nu)}}

\begin{document}

\title{Spectral dimension flow on continuum random multigraph}


\author{Georgios Giasemidis}{
  address={Rudolf Peierls Centre for Theoretical Physics, 1 Keble Road, Oxford, UK},
  email={giasemidis@physics.ox.ac.uk}
}

\author{John F. Wheater}{
  address={Rudolf Peierls Centre for Theoretical Physics, 1 Keble Road, Oxford, UK},
  email={j.wheater1@physics.ox.ac.uk}
}

\author{Stefan Zohren}{
  address={Department of Physics, 
  PUC-Rio, R. Marqu\^es de S\~ao Vicente 225, Rio de Janeiro, Brazil},
 altaddress={Rudolf Peierls Centre for Theoretical Physics, 1 Keble Road, Oxford, UK} ,
email={zohren@fis.puc-rio.br}}

\begin{abstract}
We review a recently introduced effective graph approximation of causal dynamical triangulations (CDT), the multigraph ensemble. We argue that it is well suited for analytical computations and that it captures the physical degrees of freedom which are important for the reduction of the spectral dimension as observed in numerical simulations of CDT. In addition multigraph models allow us to study the relationship between the spectral dimension and the Hausdorff dimension, thus establishing a link to other approaches to quantum gravity.
\end{abstract}

\maketitle


\section{Introduction}
The causal dynamical triangulation (CDT) approach to quantum gravity (see \cite{Ambjorn:1998xu} and \cite{Ambjorn:2012jv} for a recent review) is a non-perturbative approach which employs lattice methods; the path integral is approximated as a sum over discretised (triangulated) geometries (i.e. dynamical triangulations) with a time foliation structure (the causal assumption). Computer simulations carried out in four-dimensional CDT revealed an intriguing result \cite{Ambjorn:2005db}: a reduction of the spectral dimension from 4 at large scales to 2 at small scales where quantum effects are important. Thereafter other approaches to quantum gravity also observed this phenomenon \cite{Lauscher:2005qz, Horava:2009if, Modesto:2009kq, Modesto:2009qc}. 
However studying higher-dimensional CDT and its continuum limit is a difficult task and little progress has been made to analytically explain the results of the numerical simulations from the graph point of view. Certain fractal properties of two-dimensional (C)DTs, e.g. dimensionality, were studied with random walks on random geometries  \cite{Ambjorn:1997} and graphs \cite{Durhuus:2009sm}. We pursue these ideas further by developing a formalism to study random walks on multigraphs and arguing why the latter serves as realistic model to describe dimensional reduction in CDT.

The main notion of dimensionality we use is the {\it spectral dimension} $\dS$. It is related to a diffusion process (or random walk) on a geometric object defined by
$p(t) \sim t^{-\dS/2}$ as $t \to \infty$, 
where $p(t)$ is the probability that a simple random walker returns to its origin after time $t$. An equivalent and sometimes more efficient way to study the spectral dimension is through the generating function of return probabilities defined as 
$Q(x) = \sum _{t=0}^{\infty} p(t) (1-x)^{t/2} \sim x^{-1+\dS/2}$
as $x \to 0$
provided it diverges in this limit. Divergence of $Q(x)$ implies that $\dS \leq2 $ and the random walk is recurrent, i.e. the probability to return to the origin is one. For non-recurrent (transient) random walks there is a finite probability that the random walker escapes to infinity, in this case $Q(x)$ is finite and the spectral dimension is defined through the first diverging derivative of $Q(x)$, i.e.\ 
$Q^{(k)}(x) \sim x^{-(k+1)+\dS/2}$ as $x \to 0$
for $2k \leq \dS < 2 (k+1)$.

A second definition of dimensionality on fractal geometries is the {\it Hausdorff dimension}, $\dH$, defined by the volume growth of a ball of radius $N$ centred on a fixed vertex, i.e.
$B_N \sim N^{\dH}$ as $N\to \infty.$
Both definitions are valid provided that the limits exist and results for both the ensemble average and almost all graphs can be found in principle. 

It is known \cite{Coulhon:2000} that for fixed graphs the Hausdorff and spectral dimensions (under certain conditions) satisfy the inequalities
\beq \label{dh_ds_inequality}
\dH \geq \dS \geq \frac{2 \dH}{1+\dH}
\eeq
which are also true for some random graphs. We will present instructive examples of such random graphs and further comment on this relation in the next sections. 

\section{the multigraph approximation}

\begin{figure} \label{fig1}
  \includegraphics[scale=0.3]{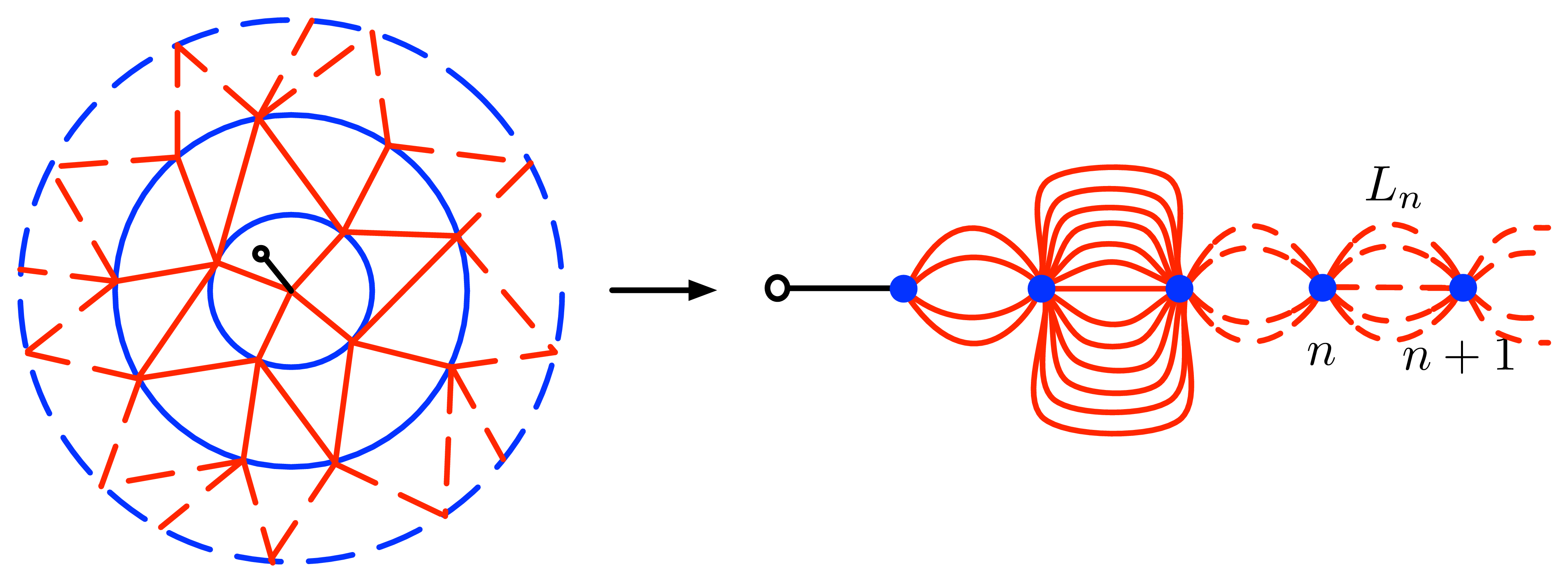}
  \caption{Illustration of how to obtain a multigraph from a causal triangulation.
  }
\end{figure}
The time foliation of $(d+1)$-dimensional causal triangulations implies a topology $I \times \Sigma^{d}$, where $d$ is the spatial dimension. The triangulation is made of triangles and their higher-dimensional  analogues ($(d+1)$-simplices) which connect the $d$-dimensional spatial hyper-surfaces. A multigraph $M$ is defined by introducing a mapping which acts on a rooted infinite causal triangulation $C$ by collapsing all space-like edges at a fixed distance $k$, $k \geq1$ from the root and identifying all vertices at this distance $k$. In the resulting multigraph a vertex $k$ has neighbours $k\pm1$, except the vertex $0$ (the root $r$) which has $1$ as a neighbour, and there are $L_k(M) \geq 1$ (time-like) edges connecting $k$ and $k+1$ (see Figure \ref{fig1}). A random walker at vertex $k$ moves to $k+1$ with probability $p_k(M) = L_k/(L_{k-1}+L_k)$ and to $k-1$ with probability $1- p_k(M)$. Note that the walker leaves the root to vertex 1 with probability one.

Denote $\eta _{M_k} \equiv Q_{M_k}/L_k$, where $M_k$ is the multigraph obtained from $M$ by removing the first $k$ vertices and the edges attached to them and relabelling the remaining graph. Then the generating function $Q_{M_k}$ follows the recursion relation \cite{Durhuus:2009sm}
\beq \label{rec_rel}
\eta _{M_k} (x) = \eta _{M_{k+1}} (x) + \frac{1}{L_k} - x L_k \eta _{M_k} (x)  \eta _{M_{k+1}} (x).
\eeq

\section{The two-dimensional model}
\label{two_dim}
A two dimensional model of causal triangulations, the uniform infinite causal triangulation (UICT), was studied in \cite{Durhuus:2009sm, Giasemidis:2012rf}. Due to a bijection between causal triangulations and planar trees, the UICT - in essence a CDT at criticality - has the same measure as  the generic random tree which can be viewed as a critical Galton-Watson process (with variance ${f^U} ''(1)=2b/(1-b)$, where $b$ controls the variance of the offspring probabilities) conditioned on non-extinction. Furthermore, by construction the multigraph ensemble inherits its measure from the UICT. There is a number of analytical results which follow from UICT \cite{Durhuus:2009sm}. First, the ensemble average of the number of edges at distance $N$ from the root and the volume of a ball of radius $N$ is given by 
\bea 
\label{connectivity_avg}
\left \langle L_N \right \rangle _{\mu} &=& N {f^U}''(1) + 1, \qquad  n\geq 1, \\
\label{volume_avg}
\left \langle B_N \right \rangle _{\mu} &\equiv& \left \langle \sum_{k=0}^{N-1} L_k \right \rangle _{\mu} = \half N (N-1) {f^U}''(1) + N, \qquad N\geq 1,
\eea
which implies $\dH = 2$. Furthermore it is analytically proven in \cite{Durhuus:2009sm} that $L_N$  is bounded above by logarithmic fluctuations around the average for almost all graphs in the ensemble, i.e. for almost all graphs
\bea \label{Ln_bounded_above}
L_N \leq c N \log N
\eea
for large $N$, where $c >1$. 
In other words, the number of space-like edges at finite height $N$, $\abs{S_N}$, remains finite since $L_N = \abs{S_N} +\abs{S_{N+1}}$. Thus omitting the space-like edges in the reduced model does not affect the random walk at large times and therefore the value of the spectral dimension of the causal triangulation. This turns to be a crucial point in our arguments in next section.

To understand further features of the multigraph approximation  we introduce the notion of {\it graph resistance} $R_G$. It is defined by considering the graph as an electric network where each edge has resistance one \cite{Lyons:2011}. One distinguishes two cases: The recurrent case ($\dS \leq 2$) where a random walker ``faces'' infinite resistance to escape to infinity and returns to the root with probability 1; and the transient (or non-recurrent) case ($\dS \geq 2$) where finite resistance to infinity implies return probability strictly less than one. 

By Rayleigh's monotonicity law the resistance from the root to infinity of the two-dimensional causal triangulation $R_{CT}$ is bounded by
$R_{M} \leq R_{CT} \leq R_{{\emph tree}}$, 
where $R_{M} $ and $R_{{\emph tree}}$ are the resistance of the corresponding multigraph and tree respectively. Given that the resistance of recurrent multigraphs is infinite this inequality implies that the two-dimensional UICT is recurrent and $\dS \leq 2$ almost surely. Furthermore it  implies that the recurrent multigraph ensemble and the generic tree ensemble are two extreme cases used to bound the spectral dimension of UICT and saturate the left and right hand side of \eqref{dh_ds_inequality} respectively (note that all the above graphs have $\dH=2$ and $\dS(GT) = 4/3$ \cite{Durhuus:2009sm, Durhuus:2006vk}). It is believed that the spectral dimension of UICT is two and that thus multigraphs provide a tight bound.

In addition it was argued in \cite{Giasemidis:2012rf} the parameter $b$ introduced above is related to a CDT with an additional term in the action proportional to the absolute value of the scalar curvature $\sum _{\upsilon}\abs{R_{\upsilon}}$ with a coupling constant $-\log b$. This term allows us to describe a scale dependent spectral dimension in the recurrent multigraph model. A scale dependent spectral dimension on graphs was studied before in \cite{Atkin:2011ak} in the context of random combs. The measure of the multigraph ensemble depends on a characteristic distance $\Lambda = b^{-1}$ and the continuum limit is defined by taking the lattice spacing $a \to 0$ 
\beq \label{Q_cont_2dim}
\tilde Q(\xi, \lambda) = \lim_{a\to0} a^{1/2} \left \langle Q(x = a\xi ; \Lambda = a^{-1/2} \lambda ^{1/2}) \right \rangle _{\mu} \sim 
\begin{cases}
\xi ^{-1/2}, &\xi >> \lambda ^{-1}, \\
\lambda ^{1/2} \abs{\log \lambda \xi},  &\xi << \lambda ^{-1},
\end{cases}
\eeq
which implies $\dS^{0} = 1$ at short scales and $\dS^{\infty} = 2$ at long scales \cite{Giasemidis:2012rf}. 

It is worth mentioning that  pure two-dimensional CDT has no length scale in the action  due to the Gauss-Bonnet theorem. But as we argued the above model with arbitrary variance depending on the parameter $b$ describes CDT with a term in the action coupling to the absolute value of the curvature which re-introduces the length scale $\Lambda = b^{-1}$ (according to dimensional analysis it is proportional to the inverse bare Newton's constant). Therefore $\sqrt{\lambda}$ can be thought  as the renormalised two-dimensional gravitational constant $G^{(2)}$.

\section{the four-dimensional model} 
\label{S:four_dim} 

Unlike in two dimensions where the measure of the multigraph ensemble is obtained analytically, the situation in higher dimensions is more complicated and only numerical results are available. However, we gained important insides from the two-dimensional model. Firstly, it suggests that the multigraph approximation gives a tight upper bound for the spectral dimension of CDT. Secondly, from \eqref{Ln_bounded_above} we argued that  the diffusion is not affected by random walks a finite amount of time in the spatial hyper-surfaces. Therefore the multigraphs  captures the degrees of freedom of the CDTs which influence the spectral dimension. Thirdly, it illustrates how the spectral dimension of the multigraph ensemble depends on two exponents: the volume growth and the resistance growth. This statement is made rigorous in \cite{Giasemidis:2012rf} where it was proven in the transient case that
\beq \label{ds_dh_rho}
\dS = \frac{2 \dH}{2 + \rho}
 \eeq 
 where $\rho \geq 0$ is an exponent which controls the anomalous resistance growth. It is seen that $\dS = \dH$ requires $\rho = 0$. We note that an equivalent expression has been found in the Einstein-Hilbert and the $R^2$ truncation of the exact renormalisation group program \cite{Reuter:2011, Rechenberger:2012pm}, where 
 $\rho$ controls the power-law change of the functional form of the Laplacian under the RG flow. 

Keeping these three points in mind we adopt three assumptions for the measure $\mu(\nu)$ of the multigraph ensemble for four-dimensional CDT, which are closely related to the volume and resistance growth. 
Firstly, we assume that the expectation value of the connectivity given by \eqref{connectivity_avg} in two dimensions generalises to  
\beq \label{connectivity_ansatz}
\left \langle L_N \right \rangle _{\mu (\nu)} \simeq \nu N^{3-\epsilon} +N
\eeq
where $\nu$ is related to the inverse bare Newton's constant and  $\epsilon$ is arbitrarily small.\footnote{Assumption \eqref{connectivity_ansatz} implies tight bounds on the volume
$\underline{c}N\avg{L_N}_{\meas} \leq \avg{B(N)}_{\meas} \leq  \bar{c}N\avg{L_N}_{\meas} $.
It is in agreement with computer simulations of four-dimensional CDT where the average number of time-like edges is bounded above by $\avg {B(t)}_{\meas} \equiv  \avg{ \sum_{n=0}^{t}  L_n}_{\meas}  \leq c' t^4$. It is also a generalisation of \eqref{volume_avg}.}

The second assumption bounds from above the resistance from vertex $N$ to infinity and the connectivity at distance $N$, i.e.
\bea \label{resistance_ansatz}
R(N) &\leq & \frac{N}{ \avg{L_N}_{\meas}}  \psi _{+}(\sqrt{\nu} N^{\frac{2-\epsilon}{2}}), \\
\label{connectivity_upper_bound}
L_N &\leq& \avg{L_N}_{\meas} \psi (\sqrt{\nu} N^{\frac{2-\epsilon}{2}})
\eea 
for $N>N_0 >0 $ and almost all graphs of the ensemble, where any $\psi(x)$ is a diverging and slowly varying function  at $x=0$ and $x=\infty$. Note that \eqref{connectivity_upper_bound} is the the four-dimensional analogue of \eqref{Ln_bounded_above} where $\psi(x) = \log x$. The description of the four-dimensional model ($2 \leq \dS \leq 4$) requires transient multigraphs with finite resistance $R(N)$. From the definition given above it follows that we have to extract $d_s$ from the first derivative of the generating function, $Q'(x)$, which is diverging.

In \cite{Giasemidis:2012rf, Giasemidis:2012qk} is was shown that differentiating the recursion relation \eqref{rec_rel}, iterating it, noting that $\eta _{M_N}(0) = R(N)$ and applying the above assumptions one gets $\avg{\abs{Q_M'(x)}}_{\meas}\sim 1/(x^{\frac{\epsilon}{2}}\nu + x)$ up to slowly varying fluctuations; 
%
thus, taking the continuum limit one has 
\beq \label{Qprime_cont_4dim}
\abs{ \tilde Q'(\xi, G) }\equiv \lim _{a \rar 0 } \left(\frac{a}{G}\right) \avg{\abs{Q_M'(x=a \xi)}}_{\mu (\nu)} \sim
\begin{cases}
\xi ^{-1}, &\xi >> G^{-1},\\
\xi ^{-\epsilon /2}, &\xi << G^{-1}
\end{cases}
\eeq
with $\nu=a^{1-\frac{\epsilon}{2}}/G$. This implies $\dS^{0} = 2$ in the short walk limit (i.e. IR limit) and $\dS^{\infty} = 4-\epsilon$ in the long walk (or UV) limit. From \eqref{Qprime_cont_4dim} we observe that the characteristic scale of the multigraph is set by the bare inverse Newton's constant $\nu$. Therefore $G$ corresponds to the renormalised Newton's constant and sets a scale for the {\it duration} of the walk. However it is the square root of it, $\sqrt{G}$, which gives the {\it length extent} on the graph and which is identified with the Planck length $l_P$.

Secondly, we apply a Tauberian Theorem to $\avg{\abs{Q_M'(x)}}_{\meas}$ to obtain the average return probability 
$\avg{p_M(t)}_{\meas}  \sim 2t^{-2}  \left( (\nu+1)/(1-1/t)^{2} -1\right)^{-1}$ for large times \cite{Giasemidis:2012qk}.
Scaling time $t(a) = \lfloor \frac{\sigma}{a}\rfloor$ and $\nu(a)$ as before we define the continuum return probability density for continuous diffusion time $\sigma$ as $\tilde{P}(\sigma) \equiv\lim _{a \rar 0} \left ( \frac{a}{G} \right )^{-1} \avg{p_M(t)}_{\mu(\nu )}$ and the scale dependent spectral dimension as $D_s (\sigma) \equiv - 2 d\log \tilde{P}(\sigma) / d \sigma$ resulting in 
\beq
\tilde{P}(\sigma) \sim   \frac{2 G^2}{\sigma^2}  \frac{1}{ 1 + 2G / \sigma}, \quad\text{and}\quad  D_s (\sigma) =  4\left(1- \frac{1}{2+ \sigma/G} \right).
\eeq
The functional form of this expression is identical with the expression for the return probability for diffusion on four-dimensional CDT conjectured in \cite{Ambjorn:2005db} 
and consistent with the numerical results.

\section{Conclusions}
In this article we discuss ``radially reduced'' models of causal quantum gravity, so-called multigraph ensembles. We argue that they capture the physical degrees of freedom which describe the phenomenon of dynamical dimensional reduction and present results related to two- and four-dimensional CDT. We first study the recurrent model which corresponds to two-dimensional CDT with an extra term which couples to the absolute value of curvature. Taking the continuum limit the spectral dimension flows from 2 at large scales to 1 at short scales. Next we presented  rigorous arguments that the spectral dimension of multigraphs depends on two ingredients; the volume and resistance growth. Our assumptions depend on these two elements accompanied with the fact that the spatial hyper-surfaces remain finite. This model reproduces the dimensional reduction from 4 in the IR to 2 in the UV in a way which is compatible with the numerical results.



\begin{theacknowledgments}
GG would like to acknowledge the support of the A.G. Leventis Foundation and
the A.S. Onassis Public Benefit Foundation grant F-ZG
097/ 2010-2011. This work is supported by EPSRC grant EP/I01263X/1 and STFC grant ST/G000492/1.
\end{theacknowledgments}



\bibliographystyle{aipproc}   



\end{document}